\documentclass[aps,twocolumn,prd,reprint,superscriptaddress,nofootinbib, raggedbottom]{revtex4-2}
\usepackage{hyperref}
\hypersetup{colorlinks=true,linkcolor=purple,anchorcolor=blue,citecolor=blue, filecolor=blue,urlcolor=blue,bookmarksnumbered=true,
	pdfview=FitB
}
\usepackage{ragged2e}
\usepackage[justification=centering]{caption}
\usepackage{subcaption}
\usepackage{color}
\usepackage{xcolor}
\usepackage{amsmath}
\usepackage{ulem}
\usepackage{csquotes}
\colorlet{purple1}{blue!70!red}
\colorlet{darkred}{red!50!black}
\usepackage{graphicx}
\usepackage{psfrag}
\usepackage{color}
\usepackage{comment}
\usepackage{amssymb}
\usepackage{amsmath}
\usepackage{epstopdf}
\usepackage{epsf}
\usepackage{subcaption}
\usepackage{natbib}

\newcommand{\nslash}{\kern 0.2 em n\kern -0.50em /}
\newcommand{\kslash}{\kern 0.2 em k\kern -0.45em /}
\newcommand{\lslash}{\kern 0.2 em l\kern -0.50em /}
\newcommand{\pslash}{\kern 0.2 em p\kern -0.50em /}
\newcommand{\Sslash}{\kern 0.2 em S\kern -0.50em /}
\newcommand{\Pslash}{\kern 0.2 em P\kern -0.50em /}
\newcommand{\Dslash}{\kern 0.2 em D\kern -0.65em /\kern 0.15em}

\newcommand{\be}{\begin{eqnarray}}
\newcommand{\ee}{\end{eqnarray}}

\begin{document}

\title{Gravitational transverse momentum dependent distributions for the gluons}

\author{Kauship Saha}
\email{kauships24@iitk.ac.in}
\affiliation{Department of Physics, Indian Institute of Technology Kanpur, Kanpur-208016, India}

\author{Dipansu Mishra}
\email{dipansum25@iitk.ac.in}
\affiliation{Department of Physics, Indian Institute of Technology Kanpur, Kanpur-208016, India}

\author{Dipankar~Chakrabarti}
	\email{dipankar@iitk.ac.in} 
	\affiliation{Department of Physics, Indian Institute of Technology Kanpur, Kanpur-208016, India}  

\author{Asmita~Mukherjee}
	\email{asmita@phy.iitb.ac.in} 
	\affiliation{Department of Physics, Indian Institute of Technology Bombay, Powai, Mumbai 400076, India}

\begin{abstract}
We investigate the momentum-space structure of the gauge-invariant gluon energy-momentum tensor (EMT) for spin-$0$ and spin-$\tfrac{1}{2}$ hadrons in the light-front formalism. We parameterize the nonlocal gluon EMT in terms of so-called gluon gravitational transverse-momentum-dependent distributions (gravitational TMDs) and derive their connections to twist-2 and twist-3 gluon TMDs. Moreover, we demonstrate that particular components of the gluon EMT encode information about the average kinetic four-momentum of gluons and the internal mechanical properties of hadrons—such as pressure and shear distributions.
\end{abstract}

\date{\today}

\maketitle

\section{Introduction}

The local QCD energy-momentum tensor (EMT) captures key insights into the internal structure of hadrons in terms of their quark and gluon degrees of freedom, providing access to fundamental properties such as the spin
decomposition~\cite{Leader:2013jra,Jaffe:1989jz,Ji:1996ek,Wakamatsu:2014zza,Lorce:2021gxs}
and mass decomposition of hadrons~\cite{Ji:1994av,Ji:1995sv,Yang:2018nqn,Hatta:2018sqd,
Lorce:2017xzd,Metz:2020vxd,Lorce:2021xku}, as well as their mechanical
properties~\cite{Burkert:2023wzr,Lorce:2018egm, Lorce:2025oot}. These aspects of hadron structure are contained in the off-forward matrix elements of the local QCD EMT, which are parameterized in terms of the so-called gravitational form factors (GFFs)
~\cite{Pagels1966,Polyakov:2018zvc, Ji:2025qax}.

In particular, the GFFs contain information about how the angular momentum of quarks and gluons is distributed inside a hadron, as well as the spatial profiles of pressure and shear forces~\cite{Lorce:2018egm, Burkert:2018bqq}. Since a direct probe of the EMT would require gravitational interactions, which are extremely weak at the microscopic scale, its properties are instead accessed indirectly via its links to other partonic distributions. Specifically, the gravitational form factors (GFFs) are connected to generalized parton distributions (GPDs)~\cite{Ji:1996ek} and generalized distribution amplitudes (GDAs)~\cite{Kumano:2017lhr}. Numerous phenomenological model-based calculations have been carried out to determine hadronic GFFs and to explore their angular momentum structure, as well as the internal pressure and shear-force distributions~\cite{Chakrabarti:2020kdc,Choudhary:2022den,Sain:2025kup,Nair:2024fit,More:2023pcy,More:2021stk, Cao:2026jzm, Tanaka:2025pny, Nair:2025sfr, Nair:2024fit}.

By extending the local EMT to a nonlocal form in a gauge-invariant manner, its link to other nonperturbative partonic distributions can be naturally generalized. The off-forward nonlocal EMT can then be expressed in terms of so-called gravitational-GPDs
which provide connections to generalized parton distributions (GPDs)~\cite{Lorce:2015lna}. Similar objects were subsequently considered in Ref.~\cite{Guo:2021aik}. In the forward limit, the nonlocal EMT has
been formulated in terms of quark gravitational TMDs, which are themselves 
related to the corresponding quark TMDs~\cite{Lorce:2023zzg, Saha:2026fwe}. While the
gravitational-TMD framework is well developed for quarks, an
 analogous systematic treatment of the forward nonlocal gluon EMT has not yet been systematically established.  
 In this paper, we promote the gauge-invariant local gluon EMT (Belinfante–Rosenfeld EMT) to a nonlocal, color-gauge-invariant form, parameterize it in terms of gluon gravitational TMDs and establish their relations to the corresponding gluon TMDs~\cite{Mulders:2000sh, Boussarie:2023izj}.

The paper is organized as follows. In Sec.~\ref{Gluon gravitational transverse momentum distribution}, we formulate the gluon transverse-momentum-dependent energy-momentum tensor (EMT) and parameterize it in terms of 32 scalar functions, which we refer to as gluon gravitational transverse-momentum-dependent distributions (gravitational-TMDs). In Sec.~\ref{Connections between gluon gravitational-TMDs and gluon TMDs}, we establish their relations with the known twist-2 and twist-3 gluon TMDs. In Sec.~\ref{Mechanical properties in terms of gluon gravitonal TMDs}, we discuss mechanical observables in momentum space in terms of the gluon gravitational-TMDs. Finally, we summarize our results and conclude in Sec.~\ref{conclusion}.
\section{Gluon gravitational transverse momentum distribution}
\label{Gluon gravitational transverse momentum distribution}
In relativistic field theory, the canonical energy-momentum tensor (EMT) derived from Noether's theorem is generally neither symmetric nor gauge invariant. However, one can construct the  Belinfante-Rosenfeld energy-momentum tensor~\cite{Belinfante:1939oxq, Belinfante1940} which is both gauge invariant and symmetric. In these sections, we introduce the transverse momentum dependent gluon energy-momentum tensor using Belinfante-Rosenfeld structure and parameterize it in terms of scalar functions which are  known as the gravitational transverse momentum dependent gluon distributions.

\subsection{Transverse momentum-dependent gluon EMT}
In QCD, the Belinfante-Rosenfeld EMT for the gluons is given by 
\begin{equation}
\begin{split}
T_{Bel,g}^{\mu\nu}(r)
={}&-2\,\mathrm{Tr}\!\left(
F^{\mu\sigma}(r)
F^{\nu}{}_{\sigma}(r)
\right) \\
&+\frac{1}{2}g^{\mu\nu}
\mathrm{Tr}\!\left(
F^{\alpha\beta}(r)
F_{\alpha\beta}(r)
\right).
\end{split}
\label{eq:Bel}
\end{equation}
 where $F_{\mu \nu}=F_{\mu \nu }^aT^a$ is the field strength tensor, which relates to the gluon field by $F_{\mu \nu}=\partial_{\mu}A_{\nu}-\partial_{\nu}A_{\mu}-ig\big[A_{\mu},A_{\nu}\big]$. To define momentum space gluon EMT, we must first make the local operator Eq.~\eqref{eq:Bel} a bilocal operator in a color gauge-invariant manner by  inserting Wilson lines, i.e.,

\begin{equation}
\begin{aligned}
T_{g}^{\mu\nu[\mathcal{U},\mathcal{U}']}(r,z)
={}&-2g_{\sigma\rho}\,
\mathrm{Tr}\!\Bigl[
F^{\mu\sigma}\!\left(r-\frac{z}{2}\right)
\mathcal{U}
\\
&\hspace{0cm}\times F^{\nu\rho}\!\left(r+\frac{z}{2}\right)
\mathcal{U}'\Bigr]
\\
&\hspace{0cm}+\frac{1}{2}g^{\mu\nu}g_{\alpha\lambda}g_{\beta\tau}\,
\mathrm{Tr}\!\Bigl[
F^{\alpha\beta}\!\left(r-\frac{z}{2}\right)
\mathcal{U}\\
&\hspace{0cm}\times
F^{\lambda\tau}\!\left(r+\frac{z}{2}\right)
\mathcal{U}'\Bigr].
\end{aligned}
\label{eq:bilocal_EMT}
\end{equation}

The Wilson lines
$\mathcal{U}\equiv\mathcal{U}(r-\frac{z}{2},r+\frac{z}{2}\,|\,\eta n)$
and $\mathcal{U}'\equiv\mathcal{U}(r+\frac{z}{2},r-\frac{z}{2}\,|\,\eta' n)$
connect the points $r-\frac{z}{2}$ and $r+\frac{z}{2}$ through infinitely
extended staple-shaped paths along the lightlike direction $n$, with
$n^2=0$, where $\eta^{(')}=-1$ denotes past-pointing and $\eta^{(')}=+1$ denotes future-pointing 
directions of the corresponding Wilson lines. The above
Eq.~\eqref{eq:bilocal_EMT} defines a gauge-invariant bilocal gluon EMT
operator. To obtain its momentum-space representation, we follow the
same procedure as for the quark case~\cite{Lorce:2023zzg}.

\begin{equation}
    T_{g}^{\mu \nu [\mathcal{U},\mathcal{U'}]}(r, k)= \int \frac{d^4z}{(2\pi)^4}e^{ik.z} T^{\mu \nu [\mathcal{U},\mathcal{U'}]}_g(r, z)
    \label{eq:mom_space_opertor}
\end{equation}

Furthermore, upon integrating Eq.~\eqref{eq:mom_space_opertor} over the gluon four-momentum, we expect to recover the local Belinfante EMT of the gluon, i.e.,
\begin{equation}
\int d^4 k T_g^{\mu \nu [\mathcal{U},\mathcal{U'}]}(r,k)
= T_{\mathrm{Bel},g}^{\mu\nu}(r).
\end{equation}

Throughout this work, we adopt the light-front formalism. For a generic four-vector $a^\mu$, we write
$a^\mu=[a^+,a^-,\boldsymbol{a}_T]$, where the light-front components are defined as
$a^\pm=(a^0\pm a^3)/\sqrt{2}$ and $\boldsymbol{a}_T=(a^1,a^2)$. We can now formally define the fully unintegrated gluon EMT by taking the forward matrix element of the operator in Eq.~\eqref{eq:mom_space_opertor}
\begin{equation}
    \Theta_{g}^{\mu \nu [\mathcal{U},\mathcal{U'}]}(P,k,N,S)=
\langle P,S \,|\, 
T^{\mu\nu [\mathcal{U},\mathcal{U'}]}_{g}(0,k)
\,|\, P,S \rangle .
\label{eq:fowardmatrix}
\end{equation}

Since the orientation of the Wilson lines is unchanged by the rescaling $n \to \alpha n$ with $\alpha>0$, the correlator is insensitive to the overall normalization of $n$ and depends solely on its direction. For convenience, we therefore employ the standard choice for the lightlike vector in light-front coordinates, following Ref.~\cite{Lorce:2015lna}, 
$ n^\mu=\left[0,1,\boldsymbol{0}_{\perp}\right]$, 
together with the rescaling-invariant four-vector $N$ defined as~\cite{Lorce:2013pza}.
\begin{equation}
    N^\mu = \frac{M^2 n^\mu}{P\cdot n}.
\end{equation}

 We work in a frame in which the transverse momentum of the hadron vanishes, such that its four-momentum is given by
\begin{equation}
    P^\mu =
    \left[
    P^+,\frac{M^2}{2P^+},\boldsymbol{0}_\perp
    \right],
    \label{eq:hardon_Kinemtic}
\end{equation}
while the momentum of the struck gluon is given by 
\begin{equation}
    k^\mu =
    \left[
    xP^+,k^-,\boldsymbol{k}_\perp
    \right].
\end{equation}
where $x=k^+/P^+$ denotes the longitudinal momentum fraction carried by the gluon. Furthermore, upon integrating over the gluon momentum $k$ in Eq.~\eqref{eq:fowardmatrix}, we obtain
\begin{equation}
\begin{aligned}
\mathcal{T}_g^{\mu\nu[\mathcal{U},\mathcal{U}']}
(P,x,\mathbf{k}_{\perp},N,S)
&= \int dk^-\,\Theta_g^{\mu\nu[\mathcal{U},\mathcal{U}']}
(P,k,N,S)
\\
&= \int \frac{dz^-\,d^2\boldsymbol{z}_{\perp}}
{(2\pi)^3}\,
e^{ik\cdot z}
\\
&\quad\times
\left.
\langle P,S|
T_g^{\mu\nu[\mathcal{U},\mathcal{U}']}(0,z)
|P,S\rangle
\right|_{z^+=0}.
\end{aligned}
\label{eq:TMD-EMT}
\end{equation}

This expression is naturally interpreted as a transverse-momentum-dependent gluon EMT and encodes information about the gluon's three-dimensional momentum-space structure.

\subsection{Parameterization in terms of gravitational-TMDs}

The fully unintegrated gluon EMT under parity, hermiticity, and time-reversal invariance must satisfy the following relations:   
\begin{equation}
\begin{aligned}
\Theta^{\mu\nu}_{g}(P,k,N,S;\eta, \eta')%
&= \Theta^{\bar\mu\,\bar\nu}_{g}(\bar{P},\bar{k},\bar{N},-\bar{S};\eta, \eta'),
\\
\Theta^{\mu\nu}_{g}(P,k,N,S;\eta, \eta')
&= \left[\,\Theta^{\nu \mu}_{g}(P,k,N,S;\eta, \eta')\,\right]^*,
\\
\Theta^{\mu\nu}_{g}(P,k,N,S;\eta, \eta')
&= \left[\,\Theta^{\bar\mu\,\bar\nu}_{g}(\bar{P},\bar{k},\bar{N},\bar{S};-\eta, -\eta')\,\right]^* ,
\label{eg:symmetry}
\end{aligned}
\end{equation}
where $v^{\bar\mu} = \bar v^{\mu} = (v^{0}, -\mathbf{v}) .$

For notational simplicity, the explicit dependence on $[\mathcal{U},\mathcal{U}']$
is suppressed in the above constraint, with their respective 
information encoded in $\eta$ and $\eta'$ throughout. Following the same spirit as Ref.~\cite{Lorce:2023zzg}, we define the transverse
component of a four-vector as
$v_T^\mu = g_T^{\mu\nu}v_\nu$, where $g_T^{\mu\nu}$ denotes the transverse
metric in the subspace orthogonal to $P$ and $N$.
\begin{equation}
g_T^{\mu\nu}
= g^{\mu\nu}
- \frac{P^\mu N^\nu + P^\nu N^\mu}{M^{2}}
+ \frac{N^\mu N^\nu}{M^{2}} .
\end{equation}

Similarly, we define the transverse Levi--Civita tensor as
\begin{equation}
    \epsilon_T^{\mu\nu}
    = \frac{\epsilon^{\mu\nu\alpha\beta}\,N_\alpha P_\beta}{M^2}.
\end{equation}
where we adopt the convention $\epsilon_{0123}=1$, which implies
$\epsilon_T^{12}=1$. We also introduce the shorthand notation
$\epsilon_T^{\mu v_T}\equiv\epsilon_T^{\mu\nu}v_{T\nu}$. The covariant spin vector $S^\mu$ is split into its longitudinal and transverse components via
\begin{equation}
S^\mu = \frac{\lambda}{M} \left( P^\mu - N^\mu \right) + S_T^\mu ,
\end{equation}
where $\lambda$ specifies the longitudinal polarization and $S_T^\mu = (0, 0, \mathbf{S}_\perp)$ encapsulates the transverse light-front polarization vector.

The most general rank-2 parameterization of the gluon EMT-TMD
$\mathcal{T}_g^{\mu\nu}(P,k,N,S;\eta)$ for spin-$0$ and spin-$\frac{1}{2}$
targets can be constructed from the available tensor structures
$P^\mu$, $N^\mu$, $k_T^\mu$, $\epsilon_T^{\mu\nu}$, and $g_T^{\mu\nu}$,
subject to the symmetry constraints given in Eq.~\eqref{eg:symmetry}. We have \footnote{The remaining possible tensor structures can be omitted, as they are not independent and can be reduced using the Schouten identity, $g^{\alpha\beta}\epsilon^{\mu\nu\rho\sigma}
+ g^{\alpha\mu}\epsilon^{\nu\rho\sigma\beta}
+ g^{\alpha\nu}\epsilon^{\rho\sigma\beta\mu}
+ g^{\alpha\rho}\epsilon^{\sigma\beta\mu\nu}
+ g^{\alpha\sigma}\epsilon^{\beta\mu\nu\rho}=0$.}

\begin{widetext}
\begin{equation}
\begin{aligned}
\mathcal{T}_{g}^{\mu\nu}
= \frac{1}{P^+}\Bigg\{&
P^\mu P^\nu b_{1e}
+ N^\mu N^\nu b_{2e}
+ k_{T}^\mu k_{T}^\nu b_{3e}
+ P^{\{\mu} N^{\nu\}} b_{4e}
+ iP^{[\mu} N^{\nu]} b_{5o}
+ P^{\{\mu} k_{T}^{\nu\}} b_{6e}
+ iP^{[\mu} k_{T}^{\nu]} b_{7o}
\\
&+N^{\{\mu} k_{T}^{\nu\}} b_{8e}
+ iN^{[\mu} k_{T}^{\nu]} b_{9o}
+ M^2 g_{T}^{\mu\nu} b_{0e}
\\[0.4em]
&-\frac{\epsilon_{T}^{k_{T} S_T}}{M}\Big(
P^\mu P^\nu b^{\perp}_{1To}
+ N^\mu N^\nu b^{\perp}_{2To}
+ k_{T}^\mu k_{T}^\nu b^{\perp}_{3To}
+ P^{\{\mu} N^{\nu\}} b^{\perp}_{4To}
+ iP^{[\mu} N^{\nu]} b^{\perp}_{5Te}
\\
&\qquad
+P^{\{\mu} k_{T}^{\nu\}}b^{\perp}_{6To}
+ iP^{[\mu} k_{T}^{\nu]} b^{\perp}_{7Te}
+ N^{\{\mu} k_{T}^{\nu\}} b^{\perp}_{8To}
+ iN^{[\mu} k_{T}^{\nu]} b^{\perp}_{9Te}
+ M^2 g_{T}^{\mu\nu} b^{\perp}_{0To}
\Big)
\\[0.4em]
&-M\Big(
P^{\{\mu} \epsilon_{T}^{\nu\} S_T} b_{1To}
+ iP^{[\mu} \epsilon_{T}^{\nu] S_T} b_{2Te}
+ N^{\{\mu} \epsilon_{T}^{\nu\} S_T} b_{3To}
+ iN^{[\mu} \epsilon_{T}^{\nu] S_T} b_{4Te}
\\
&\qquad
+k_{T}^{\{\mu} \epsilon_{T}^{\nu\} S_T} b_{5To}
+ ik_{T}^{[\mu} \epsilon_{T}^{\nu] S_T} b_{6Te}
\Big)
\\[0.4em]
&-\lambda\Big(
P^{\{\mu} \epsilon_{T}^{\nu\} k_{T}} b_{1Lo}
+ iP^{[\mu} \epsilon_{T}^{\nu] k_{T}} b_{2Le}
+ N^{\{\mu} \epsilon_{T}^{\nu\} k_{T}} b_{3Lo}
+ iN^{[\mu} \epsilon_{T}^{\nu] k_{T}} b_{4Le}
\\
&\qquad
+k_{T}^{\{\mu} \epsilon_{T}^{\nu\} k_{T}} b_{5Lo}
+ ik_{T}^{[\mu} \epsilon_{T}^{\nu] k_{T}} b_{6Le}
\Big)
\Bigg\}.
\label{eq:gluon-parameterization}
\end{aligned}
\end{equation}
\end{widetext}

where $X^{\{\mu}Y^{\nu\}}=X^{\mu}Y^{\nu}+X^{\nu}Y^{\mu}$ and $X^{[\mu}Y^{\nu]}=X^{\mu}Y^{\nu}-X^{\nu}Y^{\mu}$. The above parameterization in Eq.~\eqref{eq:gluon-parameterization} can be
organized according to the polarization state of the target. In particular,
it naturally separates into the unpolarized ($U$), transversely polarized
($T$), and longitudinally polarized ($L$) contributions. \footnote{The
coefficients $b_{i}$ correspond to the unpolarized sector,
$b_{iT}^{\perp}$ and $b_{iT}$ correspond to the transversely polarized
sector, while $b_{iL}$ correspond to the longitudinally polarized sector.}
\begin{equation}
    \mathcal{T}_g^{\mu\nu}
    =
    \mathcal{T}_U^{\mu\nu}
    +
    \mathcal{T}_T^{\mu\nu}
    +
    \mathcal{T}_L^{\mu\nu}.
\end{equation}

The real-valued functions $b_{i (e,o)}(x,\boldsymbol{k}_\perp^2)$ appearing in
Eq.~\eqref{eq:gluon-parameterization} are referred to as gluon gravitational
transverse-momentum-dependent distributions. For a spin-$0$ target, the unpolarized sector is characterized by 10 gluon gravitational TMDs, denoted by $b_i$ (viz. $b_0$–$b_9$). For a spin-$\frac{1}{2}$ target, an additional 22 gluon gravitational TMDs arise. Of these, 16 are associated with transverse polarization, namely $b_{iT}^{\perp}$ (viz. $b_{0T}^{\perp}$–$b_{9T}^{\perp}$) and $b_{iT}$ (viz. $b_{1T}$–$b_{6T}$), while the remaining 6 are associated with longitudinal polarization, namely $b_{iL}$ (viz. $b_{1L}$–$b_{6L}$). The subscripts ``$e$'' and ``$o$'' denote the naive T-even and T-odd gluon gravitational TMDs, respectively. In the collinear limit, obtained by integrating over the transverse momentum
$\int d^2\boldsymbol{k}_\perp\,\mathcal{T}_g^{\mu\nu}$, only 10 independent
gluon gravitational parton distribution functions (gravitational-PDFs) remain. Interestingly, the same number of gravitational-PDFs were obtained in the quark sector~\cite{Lorce:2023zzg}. It is important to emphasize that the parameterization in
Eq.~\eqref{eq:gluon-parameterization} differs from that used for the quark
case~\cite{Lorce:2023zzg}. This difference can be traced back to the
hermiticity constraint for the gluon EMT, given in Eq.~\eqref{eg:symmetry},
under which the Lorentz indices $\mu$ and $\nu$ are interchanged, in contrast
to the corresponding hermiticity relation for the quark case, where the
Lorentz indices are not interchanged.

\section{Connections between gluon gravitational-TMDs and gluon TMDs}
\label{Connections between gluon gravitational-TMDs and gluon TMDs}
In this section, we investigate the connection between gluon gravitational TMDs and conventional gluon TMDs. A direct experimental probe of gluon gravitational TMDs would require processes sensitive to the QCD energy-momentum tensor (EMT), which are not available in the current experimental regime. Nevertheless, model-independent relations between gluon gravitational TMDs and conventional gluon TMDs can provide a pathway toward their experimental investigation through processes in which ordinary gluon TMDs are accessible. To establish such relations, we begin with the fully unintegrated gluon correlation function in the forward limit~\cite{Lorce:2013pza}

\begin{equation}
\begin{aligned}
W^{\mu\nu;\rho\sigma}
(P,k,N,S;\eta,\eta')
={}&
\frac{1}{k\cdot n}
\int\frac{d^4z}{(2\pi)^4}\,
e^{ik\cdot z}
\\
&\hspace{-7.5em}\times
\langle P,S|
2\mathrm{Tr}\!\left[
F^{\mu\nu}\!\left(-\frac{z}{2}\right)
\mathcal{U}
F^{\rho\sigma}\!\left(\frac{z}{2}\right)
\mathcal{U}'
\right]
|P,S\rangle .
\end{aligned}
\label{eq:Gluon_correlation function}
\end{equation}
Comparing Eq.~\eqref{eq:Gluon_correlation function} with Eq.~\eqref{eq:fowardmatrix}, we get 

\begin{equation}
\begin{aligned}
\Theta_{g}^{\mu \nu}(P,k,N,S;\eta,\eta')
={}&k\!\cdot\! n\Big[
-g_{\sigma\rho}W^{\mu\sigma;\nu\rho}(P,k,N,S;\eta,\eta')
\\[-0.2em]
&\hspace{-2cm}
+\frac{1}{4}\,g^{\mu\nu}g_{\alpha\lambda}g_{\beta\tau}
W^{\alpha\beta;\lambda\tau}(P,k,N,S;\eta,\eta')
\Big].
\end{aligned}
\end{equation}
Upon performing the integration over the light-front energy $k^-$, we obtain
\begin{equation}
\begin{aligned}
\mathcal{T}_{g}^{\mu \nu}(P,x,\mathbf{k}_{\perp},N,S;\eta,\eta')
={}&xP^+\Big[
-g_{\sigma\rho}\Phi^{\mu\sigma;\nu\rho}
+\frac{1}{4}\,g^{\mu\nu}
\\[-0.2em]
&\hspace{1cm}\times
g_{\alpha\lambda}g_{\beta\tau}
\Phi^{\alpha\beta;\lambda\tau}
\Big].
\end{aligned}
\label{eq:gravitionalTMD_TMD_correlator_relation}
\end{equation}
where $\Phi_{g}^{\mu \sigma; \nu \rho}$ denotes the gluon-TMD correlator, whose explicit expression is given by~\cite{Lorce:2013pza}
\begin{equation}
\begin{aligned}
\Phi^{\mu\nu;\rho\sigma}(P,x,\mathbf{k}_{\perp},N,S;\eta,\eta')
={}&\frac{1}{xP^{+}}\int
\frac{dz^-\,d^2\mathbf{z}_{\perp}}{(2\pi)^3}\,
e^{ik\cdot z}
\\[-0.2em]
&\hspace{-4.5cm}\times
\left.
\langle P,S|
2\mathrm{Tr}\!\left[
F^{\mu\nu}\!\left(-\frac{z}{2}\right)
\mathcal{U}\,
F^{\rho\sigma}\!\left(\frac{z}{2}\right)
\mathcal{U}'
\right]
|P,S\rangle
\right|_{z^+=0}.
\end{aligned}
\label{eq: gluon-TMD}
\end{equation}

The twist-2 gluon TMD correlator can be parameterized in terms of the unpolarized (U), longitudinally polarized (L), and transversely polarized (T) components, leading to~\cite{Mulders:2000sh, Xie:2026dmo}
\begin{align}
\Phi_{U}^{+i;+j}
&=\frac{1}{2}\Bigg[
-g_{T}^{ij}G(x,\mathbf{k}_{T}^{2})
+
\left(
\frac{k_{T}^{i}k_{T}^{j}}{M^{2}}
+g_{T}^{ij}\frac{\mathbf{k}_{T}^{2}}{2M^{2}}
\right)
\nonumber\\[-0.2em]
&\hspace{2cm}
\times H^{\perp}(x,\mathbf{k}_{T}^{2})
\Bigg],
\\[2mm]
\Phi_{L}^{+i;+j}
&=\frac{1}{2}\Bigg[
-i\epsilon_{T}^{ij}\lambda\,
\Delta G_{L}(x,\mathbf{k}_{T}^{2})
+
\frac{\epsilon_{T}^{k_{T}\{i}k_{T}^{j\}}}{2M^{2}}
\lambda\,
\nonumber\\[-0.2em]
&\hspace{2cm} \times
\Delta H_{L}^{\perp}(x,\mathbf{k}_{T}^{2})
\Bigg],
\\[2mm]
\Phi_{T}^{+i;+j}
&=\frac{1}{2}\Bigg[
-g_{T}^{ij}
\frac{\epsilon_{T}^{k_{T}S_{T}}}{M}G_{T}(x,\mathbf{k}_{T}^{2})
\\[-0.2em]
&\hspace{1cm}
-i\epsilon_{T}^{ij}
\frac{\boldsymbol{k}_{T}\!\cdot\!\boldsymbol{S}_{T}}{M}
\Delta G_{T}(x,\mathbf{k}_{T}^{2})
\nonumber\\[-0.2em]
&\hspace{1cm}
+
\frac{\epsilon_{T}^{k_{T}\{i}k_{T}^{j\}}}{2M^{2}}
\frac{\boldsymbol{k}_{T}\!\cdot\!\boldsymbol{S}_{T}}{M}
\Delta H_{T}^{\perp}(x,\mathbf{k}_{T}^{2})
\nonumber\\[-0.2em]
&\hspace{1cm}
+
\frac{
\epsilon_{T}^{k_{T}\{i}S_{T}^{j\}}
+
\epsilon_{T}^{S_{T}\{i}k_{T}^{j\}}
}{4M}
\\[-0.2em]
&\hspace{1cm}\left[
\Delta H_{T}(x,\mathbf{k}_{T}^{2})
-\frac{\mathbf{k}_{T}^{2}}{2M^{2}}
\Delta H_{T}^{\perp}(x,\mathbf{k}_{T}^{2})
\right]
\nonumber
\Bigg].
\label{eq:twist-2parameterization}
\end{align}
Setting $\mu=\nu=+$, in Eq.~\eqref{eq:gravitionalTMD_TMD_correlator_relation}, we get

\begin{equation}
    \mathcal{T}_{g(X)}^{++}=-xP^+g_{Ti j}\Phi^{+i;+j}_{(X)} 
\end{equation}

where (X=U,L,T) denotes the hadron polarization. Substituting the corresponding polarization components above and using Eq.~\eqref{eq:gluon-parameterization}, we obtain 

\begin{equation}
    \begin{aligned}
        b_{1e} &= xG,
&\qquad
b_{1To}^{\perp} &= -xG_{T}.
    \end{aligned}
    \label{eq:twist-2relation}
\end{equation}

Similarly, twist-3 gluon-TMD can be parameterized as~\cite{Mulders:2000sh, Xie:2026dmo}  

\begin{align}
\Phi_{U}^{+i,+-}
&=
\frac{k_{T}^{i}}{2P^+}\,
G_{3}^{\perp}(x,\mathbf{k}_{T}^{2}),
\\[2mm]
\Phi_{L}^{+i,+-}
&=
\frac{1}{2P^+}\,
i\lambda\epsilon_{T}^{k_{T}i}
\Delta G_{3L}^{\perp}(x,\mathbf{k}_{T}^{2}),
\\[2mm]
\Phi_{T}^{+i,+-}
&=
\frac{M}{2P^+}
\Bigg[
i\epsilon_{T}^{S_{T}i}
\Delta G_{3T}^{\prime}(x,\mathbf{k}_{T}^{2})
\nonumber\\[-0.2em]
&\hspace{1cm}
+i\frac{\epsilon_{T}^{k_{T}i}}{M}
\frac{\boldsymbol{k}_{T}\!\cdot\!\boldsymbol{S}_{T}}{M}
\Delta G_{3T}^{\perp}(x,\mathbf{k}_{T}^{2})
\Bigg],
\label{eq:twist-3.1}
\end{align}

and

\begin{align}
\Phi_{U}^{ij;l+}
&=
-\frac{g_{T}^{\,l[i}k_{T}^{\,j]}}{2P^+}
H_{3}^{\perp}(x,\mathbf{k}_{T}^{2}),
\\[2mm]
\Phi_{L}^{ij;l+}
&=
\frac{i}{2P^+}
\lambda\epsilon_{T}^{ij}k_{T}^{l}
\Delta H_{3L}^{\perp}(x,\mathbf{k}_{T}^{2}),
\\[2mm]
\Phi_{T}^{ij;l+}
&=
\frac{M}{2P^+}
\Bigg[
i\,\epsilon_{T}^{ij}S_{T}^{l}\,
\Delta H_{3T}^{\prime}(x,\mathbf{k}_{T}^{2})
\nonumber\\[-0.2em]
&\hspace{1cm}
+i\frac{\epsilon_{T}^{ij}k_{T}^{l}}{M}
\frac{\boldsymbol{k}_{T}\!\cdot\!\boldsymbol{S}_{T}}{M}
\Delta H_{3T}^{\perp}(x,\mathbf{k}_{T}^{2})
\Bigg].
\label{eq:twist-3.2}
\end{align}
where
$\Delta G_{3T}'=\Delta G_{3T}-\frac{\mathbf{k_{\perp}^2}}{2M^2}\Delta G_{3T}^{\perp}$
and
$\Delta H_{3T}'=\Delta H_{3T}-\frac{\mathbf{k_{\perp}^2}}{2M^2}\Delta H_{3T}^{\perp}$.
The twist-3 gluon TMDs appearing in the above parameterizations are, in
general, complex and can be decomposed as
$\mathcal{X}=\operatorname{Re}\mathcal{X}+i\operatorname{Im}\mathcal{X}$,
with the real and imaginary parts corresponding to the T-even and T-odd
gluon TMDs, respectively. Using these definitions and setting $\mu=i$ and
$\nu=+$ in Eq.~\eqref{eq:gravitionalTMD_TMD_correlator_relation}, we obtain

\begin{equation}
    \mathcal{T}_{g(X)}^{i+}=-xP^+\bigg[\Phi^{+i;+-}_{(X)}+g_{Tjl }\Phi^{ij;l+}_{(X)} \bigg].
    \label{eq:componenti+}
\end{equation}

Similarly, substituting the corresponding polarization components into Eq.~\eqref{eq:componenti+} and using Eq.~\eqref{eq:gluon-parameterization}, we obtain
\begin{widetext}
\begin{equation}
\begin{aligned}
b_{6e} &= \frac{x}{2}\left(\mathrm{Re}(G_{3}^{\perp})
+\mathrm{Re}(H_{3}^{\perp})\right),
&\qquad
b_{7o} &= -\frac{x}{2}\left(\mathrm{Im}(G_{3}^{\perp})
+\mathrm{Im}(H_{3}^{\perp})\right),
\\[2mm]
b_{7Te}^{\perp} &= -\frac{x}{2}\left(\mathrm{Re}(\Delta G_{3T}^{\perp})
-\mathrm{Re}(\Delta H_{3T}^{\perp})\right),
&\qquad
b_{6To}^{\perp} &= -\frac{x}{2}\left(\mathrm{Im}(\Delta G_{3T}^{\perp})
-\mathrm{Im}(\Delta H_{3T}^{\perp})\right),
\\[2mm]
b_{2Te} &= -\frac{x}{2}\left(\mathrm{Re}(\Delta G_{3T}^{+})
-\mathrm{Re}(\Delta H_{3T}^{+})\right),
&\qquad
b_{1To} &= -\frac{x}{2}\left(\mathrm{Im}(\Delta G_{3T}^{+})
-\mathrm{Im}(\Delta H_{3T}^{+})\right),
\\[2mm]
b_{2Le} &= -\frac{x}{2}\left(\mathrm{Re}(\Delta G_{3L}^{\perp})
-\mathrm{Re}(\Delta H_{3L}^{\perp})\right),
&\qquad
b_{1Lo} &= -\frac{x}{2}\left(\mathrm{Im}(\Delta G_{3L}^{\perp})
-\mathrm{Im}(\Delta H_{3L}^{\perp})\right).
\end{aligned}
\end{equation}
\end{widetext}

where $\Delta G_{3T}^{+ }= \Delta G_{3T}+\frac{\mathbf{k}_{\perp}^2}{2M^2}\Delta G_{3T}^{\perp }$ and $\Delta H_{3T}^{+}= \Delta H_{3T}+\frac{\mathbf{k}_{\perp}^2}{2M^2}\Delta H_{3T}^{\perp}$. The gluon gravitational-TMDs can also be related to the gluon TMDs through the
TMD parameterization adopted in Ref.~\cite{Lorce:2013pza}

\begin{widetext}
\begin{equation}
\begin{aligned}
b_{1e} &= x f_{1}^{g},
&\qquad
b_{1To}^{\perp} &= x f_{1T}^{\perp g},
\\[1mm]
b_{6e}
&= x\left(f^{\perp g}-\bar{g}^{\perp g}\right),
&
b_{7o}
&= x\left(\bar{f}^{\perp g}-g^{\perp g}\right),
\\[1mm]
b_{7Te}^{\perp}
&= -x\left(\bar{f}_{T}^{\perp g}-g_{T}^{\perp g}\right),
&
b_{6To}^{\perp}
&= -x\left(f_{T}^{\perp g}-\bar{g}_{T}^{\perp g}\right),
\\[3mm]
b_{2Te}
&= x\left[
\bar{f}_{T}^{g}+g_{T}^{g}
-\frac{\mathbf{k}_{\perp}^{2}}{2M^{2}}
\left(
\bar{f}_{T}^{\perp g}-g_{T}^{\perp g}
\right)
\right],
&
b_{1To}
&= x\left[
f_{T}^{g}+\bar{g}_{T}^{g}
-\frac{\mathbf{k}_{\perp}^{2}}{2M^{2}}
\left(
f_{T}^{\perp g}-\bar{g}_{T}^{\perp g}
\right)
\right],
\\[3mm]
b_{2Le}^{\perp}
&= x\left(
\bar{f}_{L}^{\perp g}+g_{L}^{\perp g}
\right),
&
b_{1Lo}^{\perp}
&= x\left(
f_{L}^{\perp g}+\bar{g}_{L}^{\perp g}
\right).
\end{aligned}
\label{eq:b-to-fg-relations}
\end{equation}
\end{widetext}

We
restrict the analysis to twist-3, since a complete parameterization of gluon
TMDs beyond twist-3 is, to the best of our knowledge, not currently available
in the literature.

\section{Mechanical properties in terms of gluon gravitonal TMDs}

The notion of gluon gravitational TMDs can be interpreted in terms of the gluon  kinetic four-momentum and can be employed to investigate the mechanical properties of the hadron in momentum space.

\label{Mechanical properties in terms of gluon gravitonal TMDs}
\subsection{Average longitudinal and transverse momentum}
The component $\mathcal{T}_{g}^{+\nu}$ can be used to define the   gauge-invariant average kinetic four-momentum  as

\begin{equation}
\begin{aligned}
    \langle k^{\nu}_{\mathrm{gik}}\rangle= \int dx d^{2}\mathbf{k}_{\perp}\,\mathcal{T}^{+\nu}_g 
\end{aligned}
\end{equation}

The parameterization in Eq.~\eqref{eq:gluon-parameterization} allows us to obtain the average longitudinal momentum of the gluon as 
\begin{equation}
\begin{aligned}
\langle k^{+}\rangle
&= \int dxd^{2}\mathbf{k}_{\perp}\,\mathcal{T}^{++}_g = P^+\int dxd^{2}\mathbf{k}_{\perp}\, b_{1e} \\
&= P^+\int dx d^{2}\mathbf{k}_{\perp}\, x f_{1}^{g}.
\end{aligned}
\end{equation}
We have expressed it explicitly in terms of the unpolarized gluon TMDs using Eq.~\eqref{eq:b-to-fg-relations}, and the resulting expression coincides with the quark case~\cite{Lorce:2023zzg}. It is important to stress that, although we omit the subscript “gik” in the notation for brevity, this omission is justified for the longitudinal component, since $\langle k^+\rangle=\langle k^+_{\mathrm{gik}}\rangle=\langle k^+_{\mathrm{gic}}\rangle$, as the kinetic and canonical definitions agree for this component~\cite{Lorce:2013pza}. In contrast, for the transverse components, one has $\langle k^i_{\perp \mathrm{gik}}\rangle\neq\langle k^i_{\perp \mathrm{gic}}\rangle$. Associated average kinetic transverse momentum of the gluon is thus expressed as follows (note that $k_{T}^2=-\mathbf{k}_{\perp}^2$):
\begin{equation}
\begin{aligned}
\langle k_{\perp\mathrm{gik}}^{i}\rangle
&=\int dx\,d^2\mathbf{k}_{\perp}\,
\mathcal{T}^{+i}_g
\\[-0.2em]
&\hspace{-0.3cm}
=\epsilon_{T}^{iS_{T}}
\int dx\,d^2\mathbf{k_{\perp}}\,\bigg\{\frac{\mathbf{k_{\perp}^{2}}}{2M}
\bigl(
b_{6To}^{\perp}
+i b_{7Te}^{\perp}
\bigr)\\
&\quad-M\big(b_{1To}+ib_{2Te}\big)\bigg\}.
\end{aligned}
\label{eq:transverse_momentum_average}
\end{equation}
Since we integrated out the momentum components in Eq.~\eqref{eq:transverse_momentum_average}, the average kinetic transverse momentum can be related to the forward limit of the local gluon EMT parameterization~\cite{Ji:1996ek}. We thus obtain\footnote{Use the parameterization introduced in Ref. ~\cite{Ji:1996ek}, in terms of the gravitational form factor in the forward limit, to show that the average kinetic transverse momentum vanishes.}
%
\begin{equation}
\begin{aligned}
\langle k_{\perp\mathrm{gik}}^{i}\rangle
&=\int dx\,d^2\mathbf{k}_{\perp}\,
\mathcal{T}^{+i}_g
\\[-0.2em]
&\hspace{-0.3cm}
=\epsilon_{T}^{iS_{T}}
\int dx\,d^2\mathbf{k_{\perp}}\,\bigg\{
\biggl(
\frac{\mathbf{k_{\perp}^{2}}}{2M}b_{6To}^{\perp}-Mb_{1To}
\biggr)\\
&\quad+i\bigg(\frac{\mathbf{k_{\perp}^{2}}}{2M}b_{7Te}^{\perp}-Mb_{2Te}\bigg)\bigg\}=0.
\end{aligned}
\end{equation}

Because the average transverse momentum is zero\footnote{A similar vanishing result for the quark kinetic momentum was obtained in Ref.~\cite{Amor-Quiroz:2020qmw}.}, we arrive at the following two independent equations:
\begin{equation}
\begin{aligned}
\int dxd^2\mathbf{k}_{\perp}\,
\biggl(
\frac{\mathbf{k_{\perp}^{2}}}{2M}b_{6To}^{\perp}-Mb_{1To}
\biggr)
&=
\int dxd^2\mathbf{k}_{\perp}\,
\\
&\quad \hspace{-2em}\times
x\left(f^{ g}_T+\bar{g}^{ g}_T\right)
=0 ,
\label{eq:new_relation_1}
\end{aligned}
\end{equation}
and,
\begin{equation}
\begin{aligned}
\int dxd^2\mathbf{k}_{\perp}\,
\bigg(\frac{\mathbf{k_{\perp}^{2}}}{2M}b_{7Te}^{\perp}-Mb_{2Te}\bigg)\bigg\}
&=
\int dxd^2\mathbf{k}_{\perp}\,
\\
&\quad \hspace{-2em}\times
x\left(\bar{f}^{g}_T+g^{ g}_T\right)
=0 .
\end{aligned}
\label{eq:New_TMDs_relations2}
\end{equation}
Thus, we obtain vanishing second Mellin-moment relations for the twist-3 TMDs, which arise as a direct consequence of the vanishing kinetic average transverse momentum. A corresponding analogous relation is likewise obtained for the twist-3 vector quark TMD~\cite{Lorce:2015lna}.

\subsection{Pressure and shear distribution in momentum space}

In analogy with the momentum-space formulation used for the quark sector~\cite{Lorce:2023zzg}, we express the transverse gauge-invariant kinetic gluon EMT $\mathcal{T}_g^{ij}$ as
\begin{equation}
\begin{aligned}
\mathcal{T}_g^{ij}
&=
-g_T^{ij}\sigma
+\left(
\frac{1}{2}g_T^{ij}
-\frac{k_T^i k_T^j}{k_T^2}
\right)\Pi
+\frac{k_T^i\epsilon_T^{j k_T}
+k_T^j\epsilon_T^{i k_T}}{2k_T^2}\Pi^S
\\
&\hspace{5cm}+i\epsilon_T^{ij}\Pi^A .
\end{aligned}
\label{eq:pressure_parameterization}
\end{equation}
The first two tensor structures are $T$-even and correspond to those found in the
quark case~\cite{Lorce:2023zzg}, where $\sigma$ and $\Pi$ represent the
isotropic pressure and the shear force, respectively. The third structure,
$\Pi^S$, encodes a spin-dependent $T$-odd shear distribution. In contrast,
the $\Pi^A$ term is $T$-even for the gluon EMT, while the analogous structure
in the quark case is $T$-odd, highlighting the distinct hermiticity and
time-reversal properties given in Eq.~\eqref{eg:symmetry}. Using Eq.~\eqref{eq:gluon-parameterization}, we express $\mathcal{T}_g^{ij}$ in terms of the gluon gravitational TMDs. Comparing the resulting expression with Eq.~\eqref{eq:pressure_parameterization}, we obtain
%

%
\begin{equation}
\begin{aligned}
\sigma
&=-\frac{1}{2P^+}\bigg\{
    k_{T}^{2}b_{3e}
    +2M^2b_{0e}
    -M\epsilon_{T}^{k_{T}S_{T}}\\[-0.2em]
&\hspace{3cm}
      \times \bigg(\frac{k_{T}^2}{M^2}b_{3To}^{\perp}+2b_{5To}+2b_{0T}^{\perp}\bigg)
\bigg\}, \\[0.5em]
\Pi
&=-\frac{1}{P^+}\bigg\{
    k_{T}^{2}b_{3e}
    -M\epsilon_{T}^{k_{T}S_{T}}
      \bigg(\frac{k_{T}^2}{M^2}b_{3To}^{\perp}+2b_{5To}\bigg)
\bigg\}, \\[0.5em]
\Pi^{S}
&=-\frac{2}{P^+}\bigg\{
\lambda k_{T}^2b_{5Lo}
+(k_{T}\cdot S_{T})\,
Mb_{5To}
\bigg\},  \\[0.5em]
\Pi^{A}
&=\frac{1}{P^+}\bigg\{\lambda k_{T}^2b_{6Le}
+(k_{T}\cdot S_{T})\,
M b_{6Te}\bigg\}
\end{aligned}
\end{equation}
Consequently, the pressure and shear distributions in momentum space can be written directly in terms of the gluon gravitational TMDs.

\section{Summary}
\label{conclusion}
In this work, we begin by studying the gauge-invariant gluon energy-momentum tensor (EMT) in momentum space for spin-$0$ and spin-$\frac{1}{2}$ hadrons. We parameterized it in terms of scalar functions, which we refer to as gluon gravitational transverse-momentum distributions. For a spin-$0$ hadron, there are 10 independent gluon gravitational TMDs, while for a spin-$\frac{1}{2}$ hadron, there are an additional 22 independent gluon gravitational TMDs. We also investigate the connections between gluon gravitational TMDs and conventional gluon TMDs. In this work, we focus on the connections involving twist-2 and twist-3 gluon TMDs. Beyond twist-3, the corresponding parameterization, to the best of our knowledge, is not currently available in the literature, although their connections with gravitational TMDs can, in principle, also be established.

Furthermore, we find that these gravitational TMDs can be interpreted in terms of average gluon kinetic four-momentum. This interpretation leads to new, second Mellin-moment relations among twist-3 gluon TMDs, as given in Eqs.~\eqref{eq:new_relation_1} and~\eqref{eq:New_TMDs_relations2}. In addition, mechanical properties in momentum space, such as pressure and shear-force distributions, can be extracted from these gluon gravitational TMDs. In particular, we find that a type of shear force that depends on the target spin is associated with initial- and final-state interactions. At present, only a limited number of gluon gravitational-TMDs  can be indirectly accessed through their connections to gluon TMDs. Gluon TMDs play a central role in the study of hadron structure and are expected to be a major focus of future Electron-Ion Collider (EIC) programs in the United States and China~\cite{AbdulKhalek:2021gbh, AbdulKhalek:2022hcn, Abir:2023fpo}. Measurements of gluon TMD-sensitive observables at these facilities may therefore provide indirect constraints on the gluon gravitational structure.

Our findings provide further motivation for the theoretical study and experimental investigation of higher-twist gluon transverse-momentum distributions. It would also be valuable to explore these gluon gravitational transverse-momentum distributions using nonperturbative approaches, such as lattice QCD and phenomenological model calculations.

\bibliography{ref.bib}

\end{document}